# Electrical Characterization of hexagonal SiGe

I.Bollier[1], F. Balduini[1], M. Sousa[1], M. Vettori[2], W.H.J. Peeters[2], E.P.A.M. Bakkers[2], and H. Schmid[1]

[1] IBM Research Europe – Zürich, Rueschlikon, Switzerland

[2] Department of Applied Physics, Eindhoven University of Technology, Eindhoven, The Netherlands

We report electrical measurements on hexagonal silicon-germanium (hex-SiGe), a group IV alloy with direct bandgap. Electrical contacts are formed by metal alloying and doping is achieved using ion implantation. The metastable hex-SiGe phase is successfully recovered after implantation by solid phase recrystallization. Independent of the metal used, contacts on n-type resulted in Schottky barriers due to Fermi level pinning of hex-SiGe. Overall, this constitutes a first step towards use of hex-SiGe for optoelectronic applications.

Recent advances in material synthesis have resulted in the epitaxial fabrication of hexagonal-phase $Si_{1-x}Ge_x$ alloys [1] [2]. Compared to the abundant cubic phase, the symmetry change of the crystal structure leads to band folding from the L to the Γ-point [3], introducing a direct bandgap for $x > 0.65$, with an emission wavelength tunable from 1.8 µm to 3.4 µm [4] [5]. The favorable thermal stability of the metastable alloy and the CMOS compatibility of its constituent makes hex-SiGe a promising material for light source and detectors in photonic integrated circuits.

While detailed theoretical work [4] [6] [7] and optical measurements [5] [8] [9] [10] [11] [12] are available, electrical investigations are so far lacking. Here, we aim to fill this important gap and present the first experimental electrical characterization of hex-$Si_{20}Ge_{80}$.

Hex-SiGe samples are available as core-shell GaAs/$Si_{20}Ge_{80}$ nanowires (NW) from two batches referred to as Sample A and Sample B, fabricated approximately one year apart. Both NW samples have a VLS-grown GaAs core and an epitaxial $Si_{20}Ge_{80}$ shell [5]. The hex-SiGe NWs are assumed to be unintentionally n-doped by As during the growth process, based on photoluminescence (PL) studies [12]. Sample A features NW with length of 6.5–8 µm, GaAs core diameter of 100 nm, and a SiGe shell thickness of 85 nm. Sample B NW length are 8–9.5 µm with a core diameter 150–180 nm, and a 65 nm thick shell. Both samples have similar hex-SiGe cross-sectional areas.

NWs were mechanically transferred to Si/$SiO_2$ substrates with tungsten alignment markers. Electron beam lithography was used to define contact patterns. To yield continuous metallization across the hexagonal NW geometry, a planarization step using benzocyclobutene (BCB) was applied, illustrated in Figure 1 (b). Contacts to hex-SiGe were formed after surface oxide removal using buffered HF and immediate e-beam evaporation of metal stacks. A representative device with four metal contacts is shown in Figure 1 (a).

The IV-curve for different metal contacts (Al, Ni, Ti, and Pt) as deposited on Sample A are shown in Figure 1(d). Regardless of the metal used, the IV-curve exhibits non-linear and nearly symmetric behavior. This is indicative of the formation of a significant Schottky barrier at the two metal/semiconductor interfaces, effectively leading to a back-to-back Schottky diode configuration. The observed asymmetry within a single device indicates that even small process induced variations at the metal\semiconductor interface can impact the Schottky barrier height (SBH). Theoretical models enable the calculation of SBH for back-to-back Schottky diodes [13] [14]. Fitting the experimental data using this model lead to SBH of 0.29±0.02 eV, independent of the contact metal, and ideality factors ranging from 1.17±0.04 for Pt to 1.21±0.06 for Ni. These models typically neglect the voltage drop across the semiconductor [15] [16]. As a result, extracting the fitting parameters from the measured devices is unreliable, and their interpretation should be treated with caution. Overall, no correlation between the SBH and the metal work function was found, supporting a robust Fermi level pinning (FLP) at the semiconductor interface.

To modify the Schottky barrier (SB), each metal-semiconductor configuration was subjected to contact annealing. The results for Ni/hex-SiGe are shown in Figure 1(e) and reveal a change from non-linear to linear behavior above 250˚C and a further decrease in resistance at higher temperatures. Figure 1(c) shows a STEM image of the crystalline hex-SiGe and polycrystalline NiSiGe interface. The annealing of Ni to 250˚C for 30s creates a 25 nm thick $Ni_x(Si_yGe_{1-y})_{1-x}$ layer at the interface [17], showing a well controllable alloying, ideal for the present core-shell geometry, avoiding direct contact to the GaAs core.

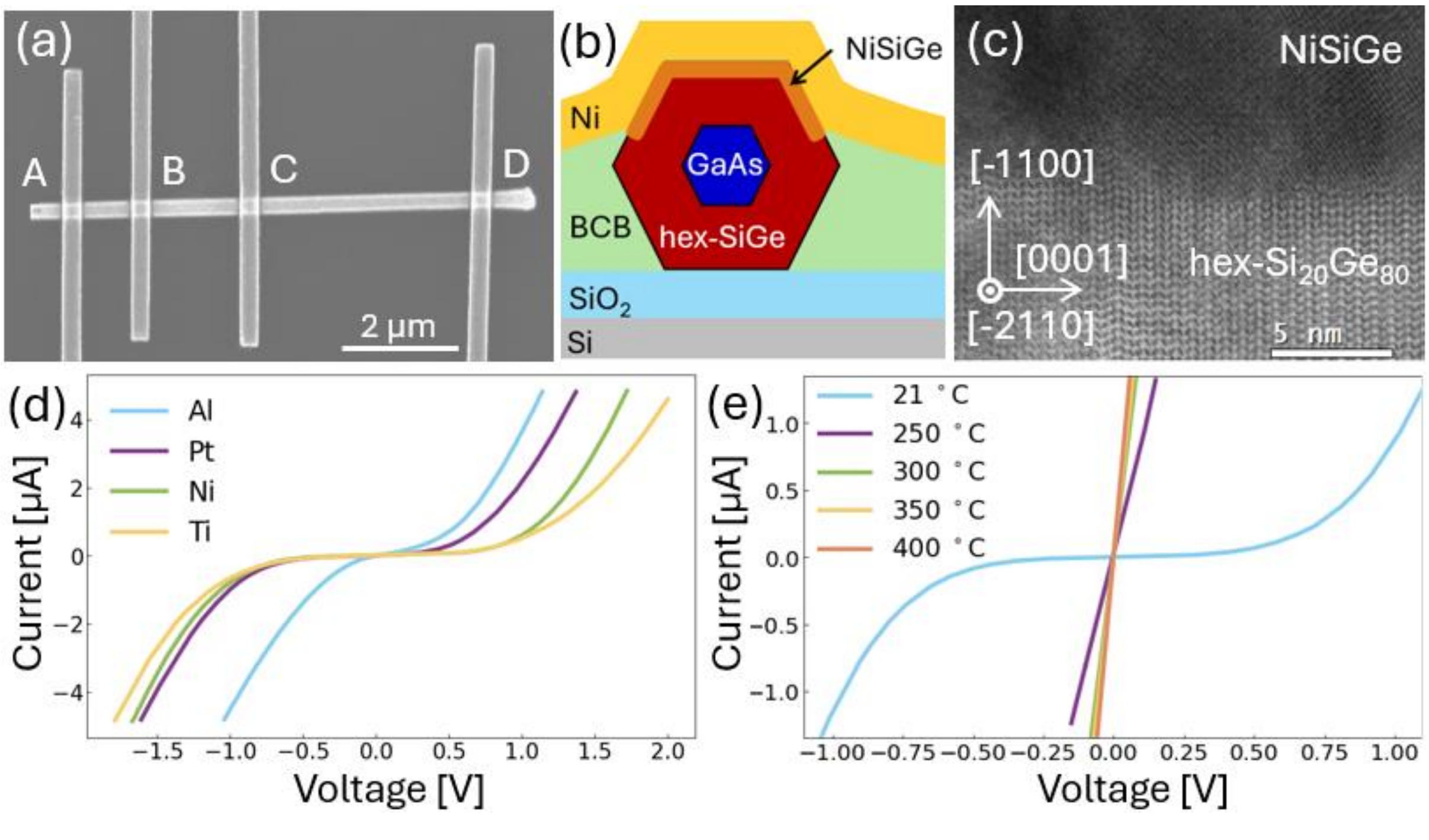

*Figure 1 Metal contacts to hexagonal core/shell GaAs/SiGe nanowire and IV-measurements. (a) Representative SEM image of a nanowire with four contacts (A, B, C, and D). (b) Schematic cross-section of a core/shell hex-GaAs/SiGe nanowire with BCB planarization layer and alloyed NiSiGe contacts. (c) STEM image of the NiSiGe/hex-SiGe interface after annealing at 250 C. (d) Sample A shows Schottky characteristics independent of metal used (not annealed). (e) Sample A with Ni contacts after indicated annealing for 30s. Contacts turn ohmic upon annealing.*

In contrast, Sample B shows linear behavior for all metal contacts even without annealing. Transfer length measurement (TLM) is used for both samples to calculate the contact resistivity and the transfer length. The results for different annealing temperatures are summarized in Table 1 (more detailed in Supplemental Material 1). The resistivity is calculated using the hex-SiGe cross-sectional area and assuming uniform conduction. Since the current path might deviate from the ideal case, the resistance per length $R_{SiGe}/L$ is additionally provided. The resistance of Sample A is about 3 times higher compared to Sample B, and a decrease of the contact resistivity with alloy formation is observed, as expected. The last column shows the percentage of functional contacts that show ohmic behavior. Annealing turns SB to ohmic and increases contact yield, while extended anneals can lead to contact failure. Ti contacts exhibit an unexpectedly high contact resistance and large transfer length, suggesting the presence of interfacial oxides. Annealed Al contacts showed unreliable results, due Al diffusion from the fast exchange reaction with SiGe [18] [19], which can result in metallic shorts between contacts (Supplemental Material 2).

*Table 1. Electrical characterization of the NW for Sample A and B using different metal contacts and are annealed at temperatures from 250 to 400 ˚C. $L_T$ indicates the transfer length, $\rho_C$ the contact resistivity, $\rho_{SiGe}$ the resistivity of the SiGe shell. The resistance per length R/L and percentege of ohmic contacts are also provided. * of working contacts.*

| Sample | Metal | Temp. | $L_T$ [µm] | $\rho_C$ ($10^{-6}$) [Ωcm$^2$] | $\rho_{SiGe}$ ($10^{-2}$) [Ωcm] | $R_{SiGe}/L$ [kΩ/µm] | Ohmic [%]* |
|---|---|---|---|---|---|---|---|
| *A* | Ni | 250 ˚C | 0.53±0.25 | 12.5±11.6 | 6.14±0.74 | 11.4±1.37 | 40 |
| | | 300 ˚C | 0.21±0.06 | 1.62±0.87 | 5.55±0.67 | 10.3±1.24 | 100 |
| | Pt | 400 ˚C | 0.37±0.15 | 5.91±3.84 | 6.25±0.81 | 11.6±1.5 | 85 |
| | Ti | 250 ˚C | 0.2±0.11 | 2.01±2.04 | 6.65±0.28 | 12.3±0.52 | 91 |
| *B* | Ni | RT | 0.75±0.35 | 8.30±6.66 | 1.70±0.20 | 3.37±0.40 | 100 |
| | | 250 ˚C | 0.52±0.14 | 3.14±1.54 | 1.54±0.19 | 3.05±0.38 | 100 |
| | | 300 ˚C | 0.42±0.17 | 2.29±1.45 | 1.64±0.60 | 3.25±1.19 | 96 |
| | Pt | RT | 1.04±0.36 | 13.6±8.7 | 1.63±0.25 | 3.23±0.50 | 98 |
| | | 400 ˚C | 0.65±0.07 | 4.38±0.78 | 1.39±0.09 | 2.75±0.18 | 100 |
| | Ti | RT | 6.28±1.45 | 512±129 | 1.82±0.45 | 3.60±0.89 | 88 |
| | | 250 ˚C | 6.07±1.27 | 494±1.04 | 1.88±0.43 | 3.72±0.85 | 86 |
| | Al | RT | 1.07±0.32 | 16.3±8.7 | 1.81±0.17 | 3.58±0.34 | 80 |

To gain further insight into the electronic properties of the hex-SiGe, carrier concentration, majority carrier type, and mobility are desired. Previous studies have demonstrated the successful application of Hall measurements to nanowire samples [20]. Attempts to perform Hall measurements on hex-SiGe samples suffered from poor signal-to-noise ratio and reproducibility (Supplemental Material 3) and were discontinued. Instead, an alternative approach was investigated based on Shubnikov-de Haas (SdH) oscillations. Here a two-contact configuration is sufficient, simplifying device fabrication compared to a Hall device. SdH magneto-transport measurements revealed periodic oscillations after background subtraction. However, the limited number of oscillations and low signal amplitudes made reliable analysis difficult (Supplemental Material 4), as observed in Te NWs [21]. As the last methodology, the applicability of field effect measurements was assessed. A top gate was fabricated on Sample B covering the entire device, and isolated from the Pt contact electrodes and the hex-SiGe by 20 nm thick $Al_2O_3$. Figure 2(a) shows a false-colored SEM image of a fabricated device. The gate electrode covers only the topmost part of the device circumferences, similarly, as illustrated in Figure 1(b), with the substrate acting as ground electrode. Figure 2(b) shows the transfer characteristics of the device, and the observed current modulation in the order of 1% of the source–drain current at the highest applied field. The small field effect is attributed to the large device cross-section and the limited gate control.

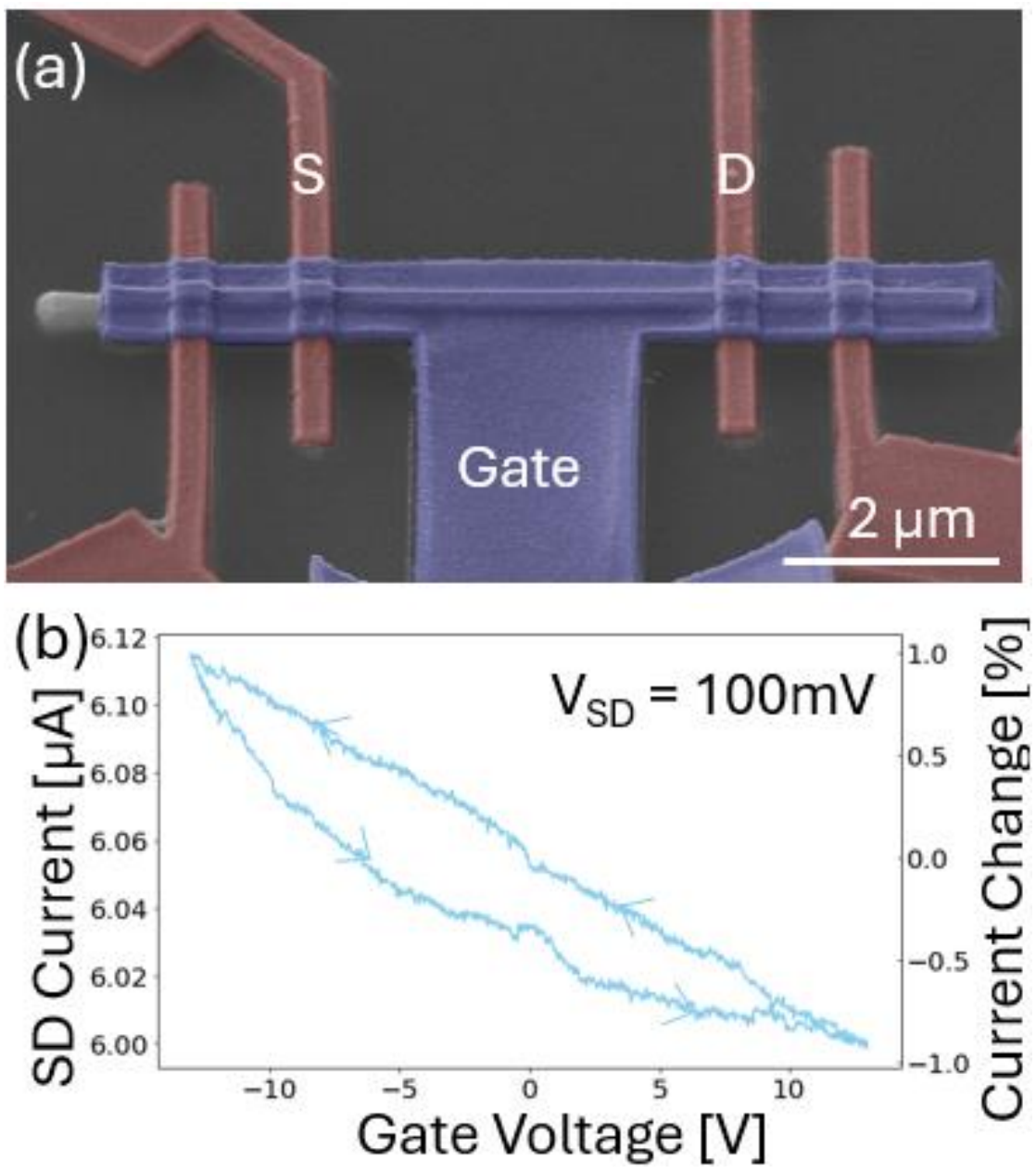

*Figure 2 Field-effect measurements on Sample B. (a) False-colored SEM image of the device with four contacts (red) with source (S) and drain (D) contacts indicated and a top gate (blue). (b) Gate induced current change exhibits a negative slop and demonstrates hole transport.*

The ungated part of the device acts as a shunt resistor that is not influenced by the gate electric field, hindering carrier depletion. Nevertheless, the observed negative slope with increasing positive gate voltage $V_G$ indicates hole-dominated (p-type) surface transport. The contribution of the hex-SiGe surface to charge transport is also observed by the strong dependence of the electronic noise on surface treatments (Supplemental Material 5). The field-effect mobility ($\mu_h$) can be calculated in the linear regime of the transfer characteristic using

$$\mu = g\,\frac{L^2}{C_{Ox}}\frac{1}{V_{SD}} \text{ where } g = \frac{\partial I_{SD}}{\partial V_G}\Big|_{V_{SD}}$$

Where g is the transconductance, L the channel length, $C_{Ox}$ the gate oxide capacitance, $V_{SD}$ the source(S)-drain(D) bias, and $I_{SD}$ is the source–drain current. Determining $C_{Ox}$ experimentally is challenging due to the small values in NW devices. Assuming the device operates in the linear regime and using an estimated capacitance of $C_{Ox} = 5{\cdot}10^{-15}$ F, based on area and a permittivity of 4, a field-effect mobility of ~ 4.3 cm$^2$/Vs at 300K is extracted from the slope of the transfer curve. The low field-effect mobility can be partially attributed to significant charge trapping and thus enhanced carrier scattering, as expected from the hysteretic transfer characteristics.

To investigate the effect of doping, ion implantation is used due its high versatility and spatial control. However, implantation introduces significant lattice damage that must be recovered by recrystallization in a high temperature post-annealing step. This poses a formidable challenge when applied to a meta-stable crystal such as hex-SiGe and exposed above amorphization threshold. Further complications arise from the core-shell heterostructure of the sample geometry. To account for the above concerns, the implantation energy was set to reach less than half the hex-SiGe shell thickness to avoid any interaction with the GaAs core, while the sufficiently large, not irradiated part of the sample acts as hexagonal lattice template for recrystallization. The recrystallization process propagates along the [-1100] direction. In contrast, a [0001] direction is expected to recrystallize into the lower energy diamond cubic phase. Figure 3(a) shows a STEM cross-section image after implantation with As at 30kV. The image reveals a nearly conformal layer of amorphous SiGe (light grey) at the upper, directly exposed surface. The transition from crystalline to amorphous is within 35 nm (Supplemental Material 6). A thin amorphous layer is also observed on the lower two inclined facettes, ascribed to recoiled ions from the substrate. Subsequently, a recrystallization step was carried out at 600 °C for 30s, low enough to avoid undesired core-shell interdiffusion. Figure 3(b) shows a STEM cross-section along the c-axis of the recrystallized sample, where only the top left side was implanted while the right was protected by a mask. The implanted region can be identified by a darker contrast. Figure 3(c) shows a lattice resolved HRTEM image of the recrystallized area, clearly demonstrating the recovery of the hexagonal lattice. Dark, vertical stripes indicate the presence of stacking defects. Whether the defects originate from clustering of impurities or are related to I3 defects [22] [23] needs further investigation.

The electrical activity of uniformly implanted and recrystallized samples was investigated by fabricating TLM structures. Importantly here, the metal contacts were not annealed to exclude possible penetration of the NiSiGe alloy through the thin doped layer to the underlying hex-SiGe.

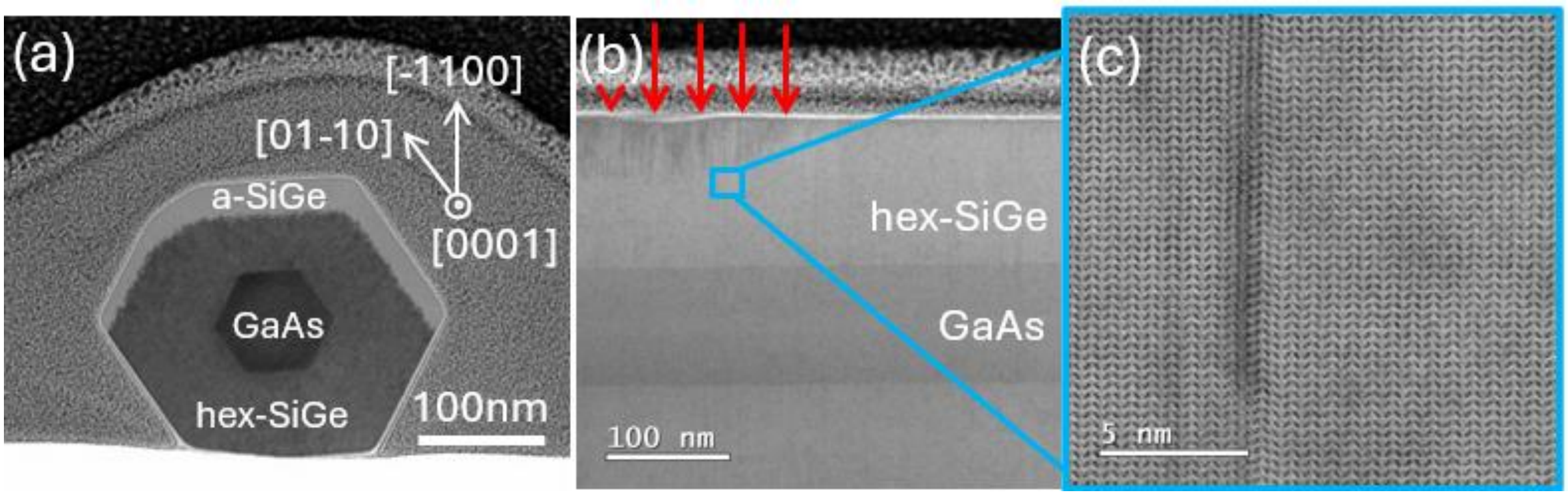

*Figure 3 Doping of hex-SiGe by ion implantation and electrical results. (a) STEM cross-section of a core/shell nanowire along the [0001]-direction after ion implantation from top, with visible amorphous region. (b) cross-section along the [11-20]-direction of the partially implanted (arrows) and recrystallized nanowire. (c) HRTEM image of the recrystallized region showing a recovered hexagonal lattice. Dark contrast reveals presence of stacking faults.*

Table 2 summarizes the results of B, Ga and As implants from Sample A and B. B and As were implanted with a dose of $8 \cdot 10^{13}$ cm$^{-2}$ and $2 \cdot 10^{14}$ cm$^{-2}$, and Ga was implanted using a Focused Ion Beam (FIB). The arsenic, differs from the others, as only a Schottky barrier characteristic was observed, hindering further analysis. Conversely, all p-type doped samples show ohmic contacts. Note that the current path in the devices is not well defined because of the complex geometry. To allow for a comparison, the resistivity calculation was simplified using the entire hex-SiGe cross sectional area. Despite the spread in the data, it can be concluded that B implants have a smaller impact on the electrical transport compared to the Ga implants, which indicates a higher Ga-dose from the FIB. Overall, despite the uncertainties of the absolute values, the data show that the implantation and recrystallization lead to electrically active impurities in hex-SiGe.

*Table 2 Effect of doping on the resistivity $\rho_{SiGe}$ and contact resistance $\rho_C$ of the NW from Sample A and B, having Al and Ni as metallic contacts. $L_T$ indicates the tranfer length, R/L the resistance per length and the last column idicates the percentege of ohmic contacts.*

| Sample | Dopant | Metal | $L_T$ [μm] | $\rho_C$ ($10^{-6}$) [Ωcm$^2$] | $\rho_{SiGe}$ ($10^{-2}$) [Ωcm] | $R_{SiGe}$/L [kΩ/μm] | Ohmic [%]* |
|---|---|---|---|---|---|---|---|
| *A* | B | Al | 0.37±0.16 | 6.37 ± 4.57 | 6.90 ± 2.67 | 12.77 ± 4.94 | 98 |
| | Ga | Ni | 1.58±1.24 | 11.0 ± 9.3 | 1.09 ± 1.05 | 2.009 ± 1.94 | 100 |
| | As | Ni | - | - | - | - | Schottky |
| *B* | B | Ni | 0.56±0.49 | 3.65 ± 4.94 | 1.37 ±0.3 | 2.721 ± 0.59 | 98 |

For devices, formation of junctions is essential. As the Samples are expected to be n-type doped, Sample A was partially p-doped with Ga to form a pn-junction. Ohmic contacts were achieved by annealing Ni contacts on the undoped side, as illustrated schematically in Figure 4(a) and shown in the STEM image in Figure 4(b). On the p-doped side, the contacts are expected to be ohmic after deposition. The contact properties were assessed by measuring A-B and C-D separately (Figure 4(c)) confirming ohmic behavior. The IV-curve of the implanted side (C-D) shows a reduced resistance (steeper slope) compared to the non-implanted side (A-B), confirming the presence of an electrically active Ga-doped section. Measurements between the outer contact pairs (A,B-C,D) resulted in linear resistances as well indicating the absence of a pn-junction in the current path. This indicates that the carriers flowing are holes and can be explained by a surface current effectively shunting the expected n-type character of the bulk hex-SiGe.

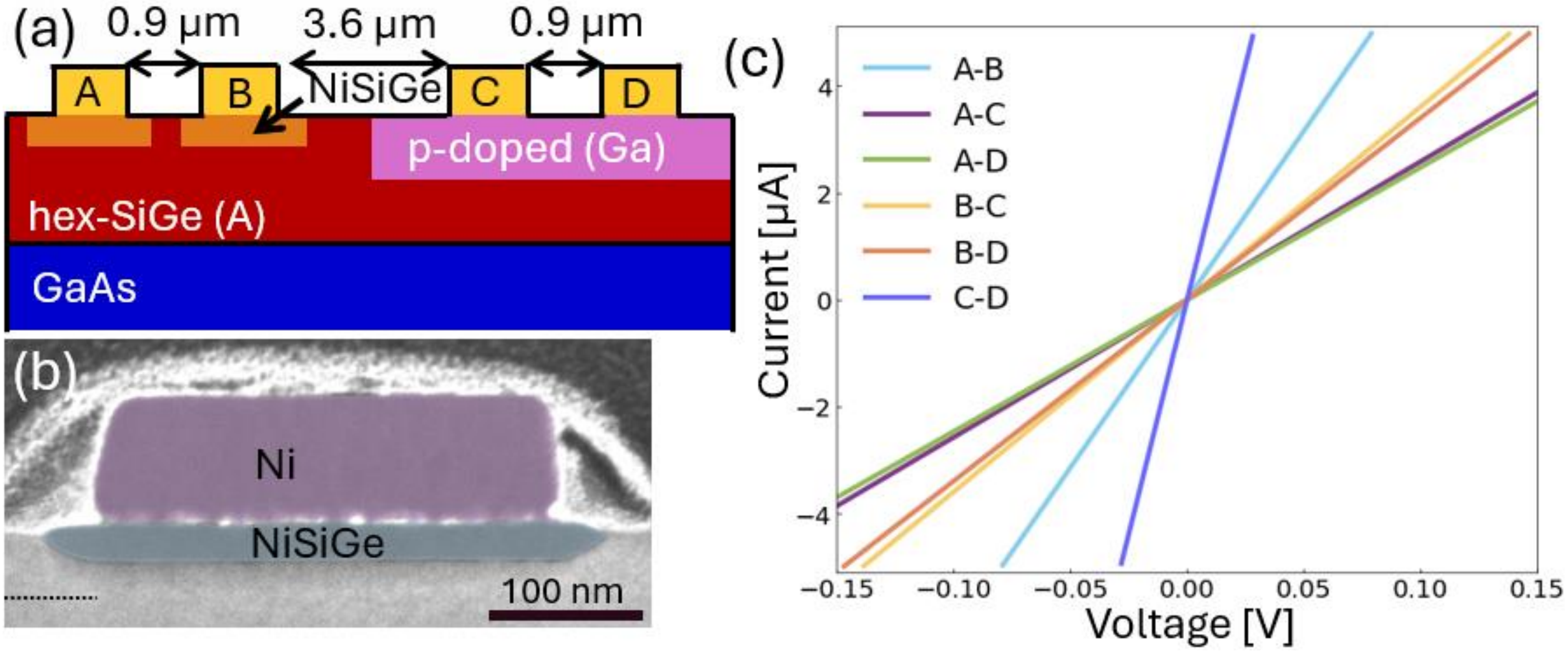

*Figure 4 Device with one-sided p-type (Ga) implanted segment. (a) Schematic of the device configuration with two annealed contacts (A and B) and two contacts on the Ga-implanted section (C and D). Only the upper half of the core-shell NW is shown. (b) False-colored STEM image of an annealed Ni contact, showing a 25nm thick NiSiGe alloy layer. (c) IV-measurements of the device shows ohmic behavior for all contact configurations.*

The second attempt to fabricate a one-sided rectifying junction is based on the properties of Sample A to form Schottky contacts. To overcome a back-to-back Schottky situation, two contacts (A, B) were deposited and annealed forming ohmic contacts, followed by the deposition of two more contacts (C, D) forming a Schottky barrier, as illustrated in Figure 5(a). Similarly to the previous device measurements, each contact pair was tested confirming ohmic contacts (A-B) and Schottky barriers (C-D) as shown in the plot in Figure 5(b). Measurements across the differently engineered contacts show symmetric behavior around zero bias. The pairs sharing the same Schottky contact (AC and BC vs. AD and BD) behave similarly, indicating that the I-V characteristics is dominated by the specific (C,D) Schottky contact. The rather symmetric behavior of the device and the deviation from a Schottky diode behavior are attributed to high junction leakage dominating the device characteristics. The origin of the leakage was probed by performing temperature dependent measurements shown in Figure 5(c). The reverse current was analyzed using the Richardson plot and the extracted SB and ideality factor plotted depending on series resistance (Supplemental Material 7). A barrier height below 200meV was extracted, corresponding to a nearly transparent (Schottky) contact, impeding efficient rectification.

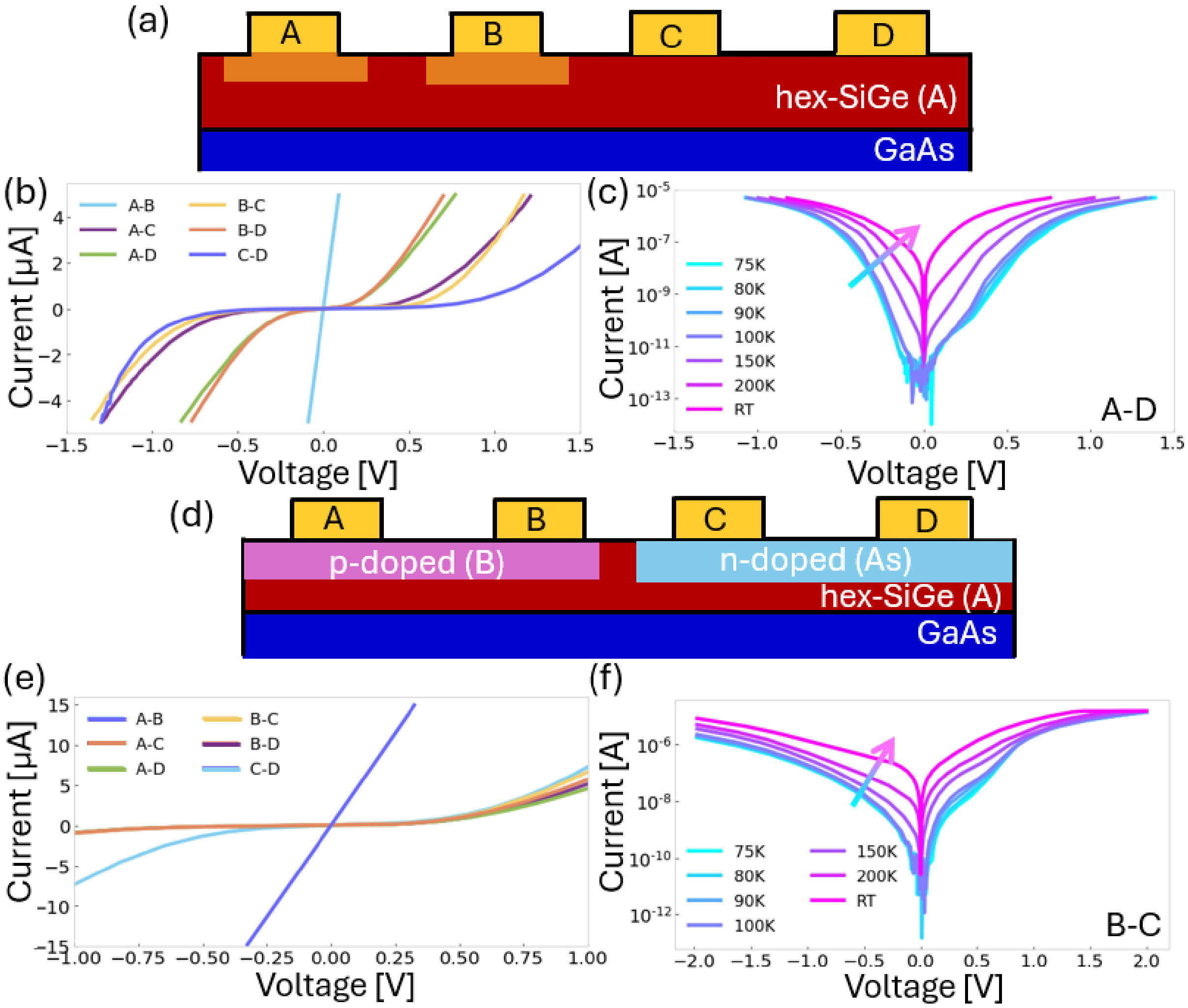

*Figure 5 Fabrication and measurements of devices containing junctions. (a) Schematic of a device, with two annealed ohmic contacts (A and B) and two Schottky contacts (C and D). (b) Symmetric IV-characteristics are observed between ohmic and Schottky contacts at room temperature. (c) Temperature dependent measurement between contact A (ohmic) and D (Schottky). (d) Schematic of a device, with two ohmic contacts (A and B) on a p-doped section and two Schottky contacts (C and D) on a n-doped section. (e) Asymmetric IV-characteristics are observed between contacts on p- and n-doped regions at room temperature. (f) Temperature dependent measurement between contact B (ohmic, p-doped region) and C (Schottky, n-doped region) showing rectification.*

The last device fabricated and illustrated in Figure 5(d), contains a boron (p-type) and an arsenic (n-type) doped segment on opposite ends forming a pn-junction. The contacts to the p-side are ohmic, while the contacts to the n-doped side form Schottky barriers. Thus, the device is a pn-junction is in series with a Schottky diode. The corresponding IV-curves in Figure 5(e) show the expected contact behavior on each side (ohmic and Schottky) while measurements across the junctions show rectifying properties. The high leakage current is an indication of a small barrier and possibly defects in the junction. Temperature dependent measurements in Figure 5(f) show a pronounced lowering of the reverse current branch and improved rectification at low temperature, a first indication of diode function in hex-SiGe.

Devices fabricated from two batches of hex-SiGe NWs (Sample A and Sample B) exhibit subtle variances in electrical resistivity and contact behavior which might originate from a different Fermi level position, possibly from a lowered n-impurity level in Sample B. Beyond the challenging sample geometry that demands conservative fabrication and analysis, a major issue is the persistent Schottky barrier observed on n-type. For hexagonal SiGe, the charge neutrality level resides deep within the valence band [7], which can lead to a hole accumulation layer at the surface, and the formation of a p-type surface channel, consistent with the findings from our gate study. Consequently, the fabrication of a one-sided Schottky diode is not possible, and junction configurations are limited to a one-sided ohmic p-type and a pinned (n-type) Schottky contact as second electrode (Fig. 5), which can be approximated as pn-diode in series with a Schottky diode. The exact nature of the device operation needs further studies including device simulations to adequately capture the underlying physics.

We demonstrated electrical contact formation on hex-SiGe using Al, Ni, Pt, Ti and contact alloying by annealing. Optimized annealing improved yield of ohmic contacts and contact resistivity. High dose implantation of B, Ga and As was performed and lead to a complete amorphization of the lattice. The metastable hex-SiGe phase was successfully recovered by solid phase recrystallization. IV-measurement showed electrical activity of the implanted dopants, and the formation of ohmic contacts on p-doped sections, whereas n-doped sections resulted in Schottky contacts independent of the contact metal. The FLP on n-type contacts hindered the measurement of pn junctions but instead formed a one-sided Schottky diode. The fabricated devices showed high reverse currents at room temperature and increasing rectification with lowering temperatures. Overall, we were able to demonstrate first promising steps towards hexagonal group-IV devices. Further work on overcoming the FLP will be of great value.

The autors have no conflict to disclose.

W.H.J.P. and M.V. carried out the growth of the NW. I.B. fabricated the devices, measured the IV-characteristics and analyzed the data. F.B. performed the magnetotransport measurements. H.S. performed the FIB cuts, and M.S. carried out the STEM analysis. H.S. supervised the project. I.B. and H.S. wrote the manuscript. All authors discussed the results and commented on the manuscript.

Supplementary Material includes extended TLM data, STEM analysis of contact and recrystallization anneals, Hall and SdH measurements, noise measurements, and Richardson plot.

The authors thank the cleanroom operations team of the BRNC and Heike Riel for supporting the work. The research leading to these results has received funding from the European Union's Horizon 2020 research and innovation program under Grant Agreement No. 964191.

# Supplemental Material

## S1: Additional TLM results

Table S1.1 shows the TLM results for Sample A, and Table S1.2 for Sample B, for different metals (Ni, Pt, Ti, and Al). These measurements require ohmic contacts, which are achieved on Sample A only after annealing. The temperature for which the contacts become ohmic depend on the contact metal and is indicated by the start of the table. Increasing the temperature further can lead to damaging on the contacts, decreasing the yield of the working devices and therefore make TLM measurements impossible. STEM studies on annealed Al contacts (Figure S2) show uncontrolled diffusion of the Al into the wire. This hinders the estimation of the separation between the contacts, leading to false TLM measurements. Therefore, these results were not further investigated.

*Table S1.1 Sample A*

| **Metal** | **Temp.** | **$L_T$ [µm]** | **$\rho_C$ [Ωcm$^2$]** | **$\rho_{SiGe}$ [Ωcm]** | **$R_{SiGe}$/L [kΩ/µm]** | **Ohmic [%]*** |
|---|---|---|---|---|---|---|
| **Ni** | RT | -[1] | -[1] | -[1] | -[1] | 0 |
| | 250 °C | 0.53±0.25 | (1.25±1.16)$10^{-5}$ | (6.14±0.74)$10^{-2}$ | 11.4±1.37 | 40 |
| | 300 °C | 0.21±0.06 | (1.62±0.87)$10^{-6}$ | (5.55±0.67)$10^{-2}$ | 10.3±1.24 | 100 |
| | 350 °C | 0.18±0.05 | (1.01±0.48)$10^{-6}$ | (4.95±0.59)$10^{-2}$ | 9.16±1.10 | 100 |
| | 400 °C | 0.41±0.26 | (5.64±6.32)$10^{-6}$ | (4.33±0.58)$10^{-2}$ | 8.02±1.07 | 90 |
| **Pt** | RT | -[1] | -[1] | -[1] | -[1] | 0 |
| | 250 °C | -[1] | -[1] | -[1] | -[1] | 0 |
| | 300 °C | -[1] | -[1] | -[1] | -[1] | 0 |
| | 350 °C | -[1] | -[1] | -[1] | -[1] | 0 |
| | 400 °C | 0.37±0.15 | (5.91±3.84)$10^{-6}$ | (6.25±0.81)$10^{-2}$ | 11.6±1.5 | 85 |
| **Ti** | RT | -[1] | -[1] | -[1] | -[1] | 0 |
| | 250 °C | 0.2±0.11 | (2.01±2.04)$10^{-6}$ | (6.65±0.28)$10^{-2}$ | 12.3±0.52 | 91 |
| | 300 °C | -[2] | -[2] | -[2] | -[2] | |
| | 350 °C | -[2] | -[2] | -[2] | -[2] | |
| | 400 °C | -[2] | -[2] | -[2] | -[2] | |
| **Al** | RT | -[1] | -[1] | -[1] | -[1] | 0 |
| | 250 °C | -[1,3] | -[1,3] | -[1,3] | -[1,3] | 0 |
| | 300 °C | -[3] | -[3] | -[3] | -[3] | 90 |
| | 350 °C | -[3] | -[3] | -[3] | -[3] | 100 |
| | 400 °C | -[3] | -[3] | -[3] | -[3] | 100 |

*Table S1.2 Sample B*

| **Metal** | **Temp.** | **$L_T$ [µm]** | **$\rho_C$ [Ωcm$^2$]** | **$\rho_{SiGe}$ [Ωcm]** | **$R_{SiGe}$/L [kΩ/µm]** | **Ohmic [%]*** |
|---|---|---|---|---|---|---|
| **Ni** | RT | 0.75±0.35 | (8.30±6.66)$10^{-6}$ | (1.70±0.20)$10^{-2}$ | 3.37±0.40 | 100 |
| | 250 °C | 0.52±0.14 | (3.14±1.54)$10^{-6}$ | (1.54±0.19)$10^{-2}$ | 3.05±0.38 | 100 |
| | 300 °C | 0.42±0.17 | (2.29±1.45)$10^{-6}$ | (1.64±0.60)$10^{-2}$ | 3.25±1.19 | 96 |
| | 350 °C | 0.62±0.27 | (3.94±3.13)$10^{-6}$ | (1.32±0.26)$10^{-2}$ | 2.61±0.51 | 95 |
| | 400 °C | -[1] | -[1] | -[1] | -[1] | 55 |
| **Pt** | RT | 1.04±0.36 | (1.36±0.87)$10^{-5}$ | (1.63±0.25)$10^{-2}$ | 3.23±0.50 | 98 |
| | 250 °C | 0.90±0.41 | (1.03±0.90)$10^{-5}$ | (1.61±0.33)$10^{-2}$ | 3.19±0.65 | 100 |
| | 300 °C | 0.95±0.30 | (1.05±0.67)$10^{-5}$ | (1.47±0.11)$10^{-2}$ | 2.91±0.23 | 100 |
| | 350 °C | 0.72±0.13 | (5.87±2.16)$10^{-6}$ | (1.50±0.08)$10^{-2}$ | 2.97±0.16 | 100 |
| | 400 °C | 0.65±0.07 | (4.38±0.78)$10^{-6}$ | (1.39±0.09)$10^{-2}$ | 2.75±0.18 | 100 |
| **Ti** | RT | 6.28±1.45 | (5.12±1.29)$10^{-4}$ | (1.82±0.45)$10^{-2}$ | 3.60±0.89 | 88 |
| | 250 °C | 6.07±1.27 | (4.94±1.04)$10^{-4}$ | (1.88±0.43)$10^{-2}$ | 3.72±0.85 | 86 |
| | 300 °C | 6.44±1.15 | (5.71±1.23)$10^{-4}$ | (1.88±0.36)$10^{-2}$ | 3.72±0.71 | 80 |
| | 350 °C | -[2] | -[2] | -[2] | -[2] | 56 |
| | 400 °C | -[2] | -[2] | -[2] | -[2] | |
| **Al** | RT | 1.07±0.32 | (1.63±0.87)$10^{-5}$ | (1.81±0.17)$10^{-2}$ | 3.58±0.34 | 80 |
| | 250 °C | -[3] | -[3] | -[3] | -[3] | |
| | 300 °C | -[3] | -[3] | -[3] | -[3] | |
| | 350 °C | -[3] | -[3] | -[3] | -[3] | |
| | 400 °C | -[3] | -[3] | -[3] | -[3] | |

*) of all the working contacts
1) Schottky contacts
2) TLM failed due to damaged contacts
3) TLM compromised due to excessive diffusion of Al

## S2: Contact annealing

Figures S2.1 Alloyed Ni contacts on hex-SiGe. (b)-(d) present EDX maps for Ni, O, and Ge of alloyed Ni contacts, also revealing an oxygen layer between the Ni and NiSiGe. During annealing, Ni diffuses through the surface oxide of the hex-SiGe, while Ge diffuses into the Ni. This interdiffusion forms a bridge between the SiGe and Ni, enhancing contact formation and reducing resistance.

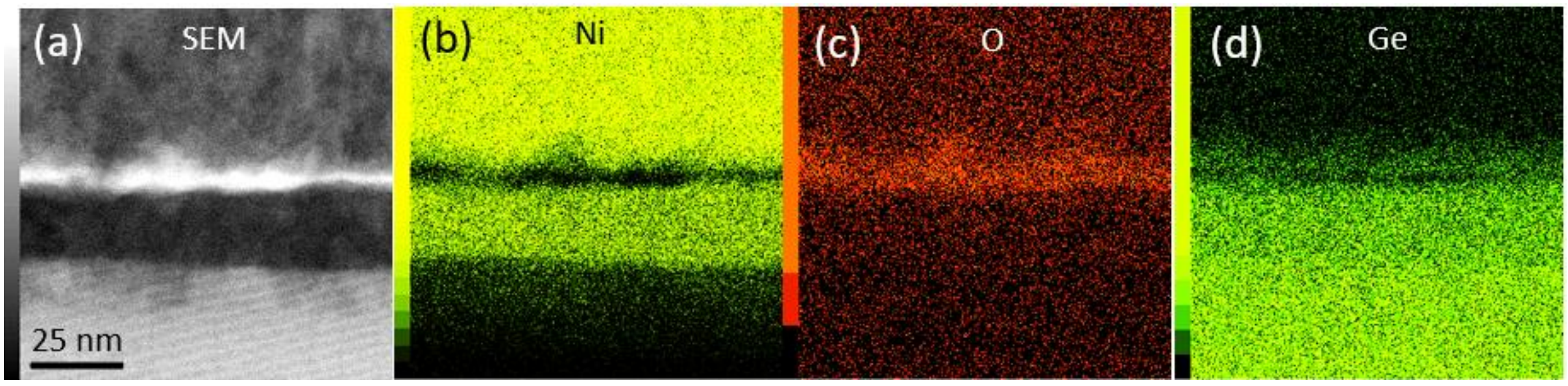

*Figure S2.1 Formation of NiSiGe alloys on hex-SiGe. (a) SEM image oft he hex-SiGe/NiSiGe/Ni interface. EXD map of the interface for (b) Ni, (c) O, and (d) Ge.*

Aluminum is a known contact metal for SiGe, but the diffusion is difficult to control. Figure S2.2 (a) shows an SEM image after annealing to 400 °C, revealing a distinct contrast change around the contact area. The corresponding EDX maps in Figures S2.2 (b) and (c) show that Al diffuses into the SiGe nanowire, completely replacing Ge in some regions. The STEM cross-section Figure 2.2(d) further illustrates this, with Al depleted contact regions and Ge largely replaced by Al. Although the diffusion appears uncontrolled, it is notably slower along the <0001> direction, resulting in a relatively well-defined interface. These observations indicate that Al is not suitable for forming well-controlled, shallow junctions required in this work.

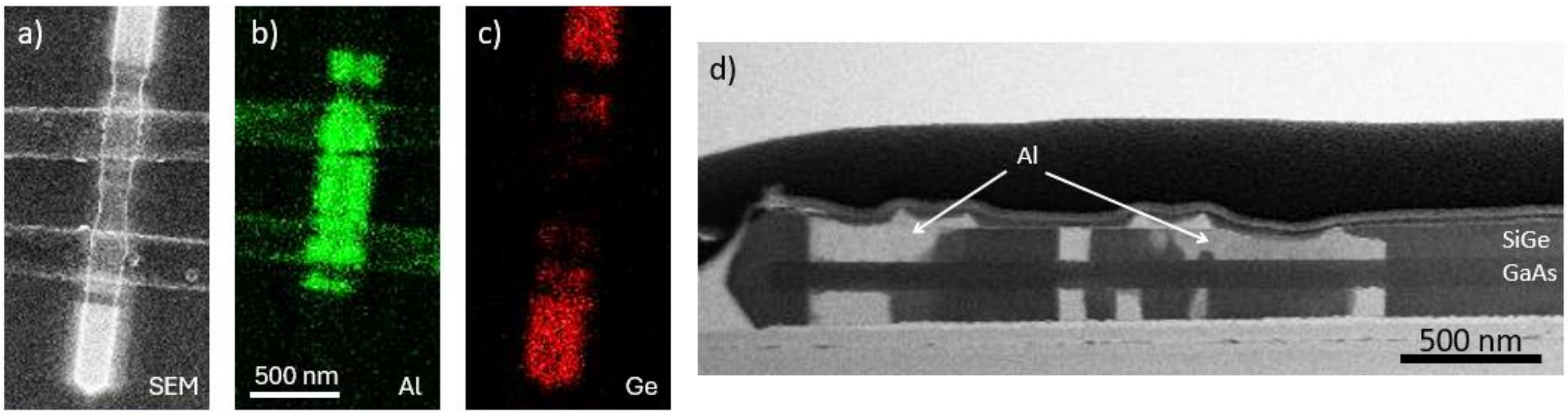

*Figure S2.2 EDX and STEM data of annealed Al contacts. a) SEM image of a nanowire with Al contacts after annealing to 400, ◦C. EDX map for b) Al and c) Ge after annealing. D) cross-section STEM image of the nanowire after annealing.*

## S3: Hall measurements

Hall effect measurements were performed on hex-SiGe nanowires (Sample B) using Ni contacts fabricated perpendicular to the nanowire axis. To measure the Hall resistivity, a magnetic field was applied perpendicular to the nanowire and the Hall resistance $R_{xy}$ was recorded over a magnetic field range of –9 to 9 T (Figure S3.1(b)). The resulting data showed a parabolic rather than linear trend, attributed to a contribution from the longitudinal resistance $R_{xx}$, caused by misaligned Hall contacts. The antisymmetrized data (Figure S3.1(c)) exhibited substantial noise. Additionally, the contact geometry was suboptimal, since the contact widths exceeded their spacing (J.G. Gluschke, J. Seidl, H. Hoe Tan, C. Jagadish, P. Caroff, A. P. Micolich, Nanoscale, Bd. 12, pp. 20317-20325, 2020) which, along with an ill-defined current path, introduced significant uncertainty. Nevertheless, assuming a cross-section thickness of 100 nm, a linear fit allowed estimation of the carrier concentration, with its temperature dependence shown in Figure S3.1(d). At 2 K the extracted carrier concentration was $n=3.52\times10^{22}\,cm^{-3}$ and the mobility $\mu=9.39\times10^{-3}\,cm^2/Vs$, obviously yielding unphysical results. While conceptually simple, the implementation of the Hall measurements on our nanowire samples proved to be not realizable.

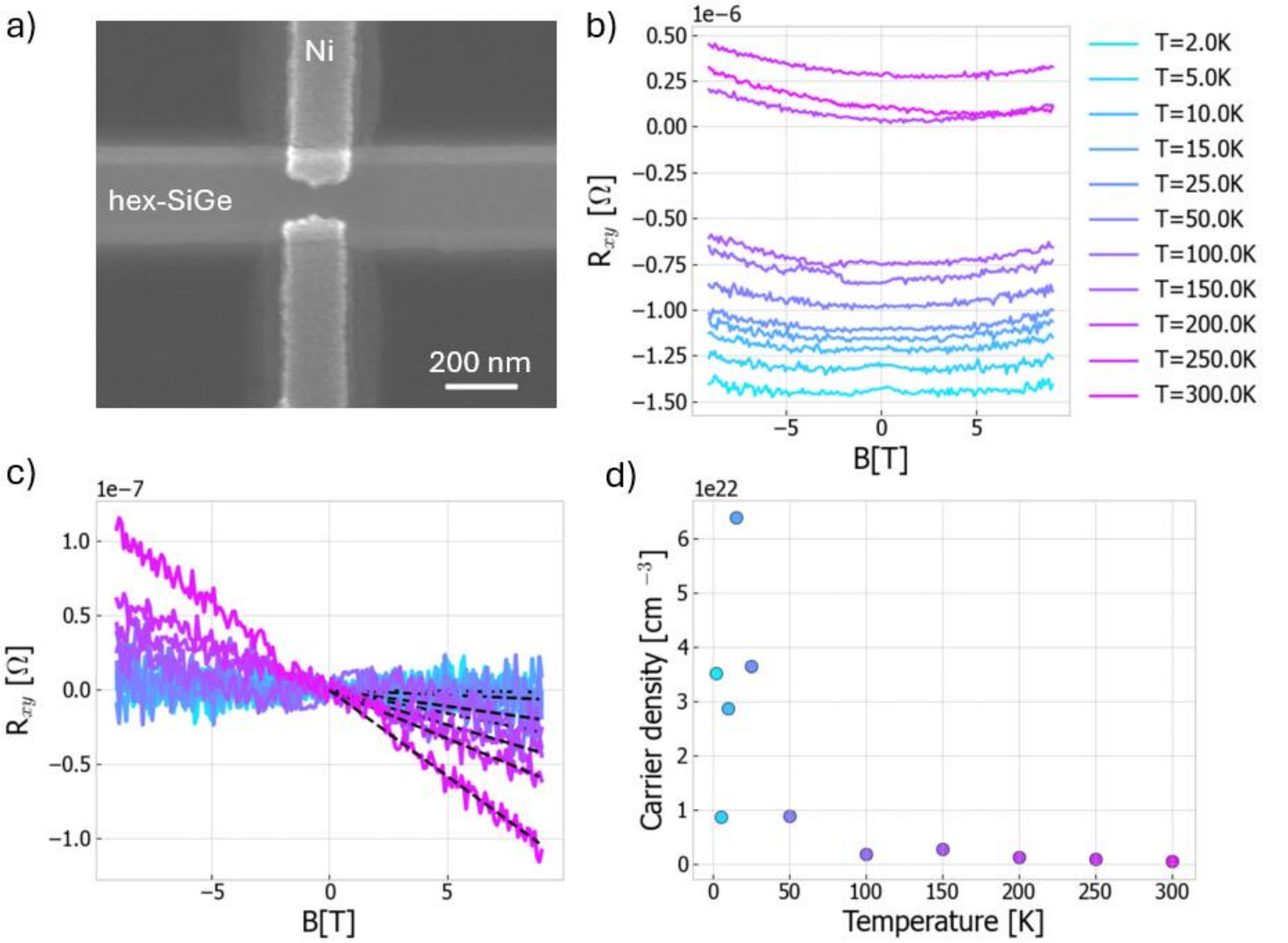

*Figure S3.1 Hall measurements a) SEM image of a hex-SiGe NW with Ni Hall contacts. b) Temperature dependence of the Hall resistances vs. magnetic field. c) Antisymmetrized Hall resistance vs. magnetic field. d) temperature dependence of the calculated carrier concentration led to unphysical values.*

## S4: Shubnikov–de Haas (SdH) oscillations

Shubnikov–de Haas (SdH) oscillations are quantum oscillations in the longitudinal resistance, enabling the use of contacted nanowires without additional processing. These oscillations were measured in a cryostat capable of applying magnetic fields perpendicular to the electric current from –9 to 9 T. The temperature-dependent magnetoresistance, representing the resistance variation with magnetic field, is shown in Figure S4.1a). Clear oscillations are visible in all the measured devices, becoming more pronounced at lower temperatures, suggesting a quantum mechanical origin. Figure S4.1b) compares the magnetoresistance for two different currents. While higher currents are generally preferred for signal enhancement, in this case, increased current led to reduced magnetoresistance and suppressed oscillations due to self-heating. To minimize this effect, a current of 200 nA was used for subsequent measurements.

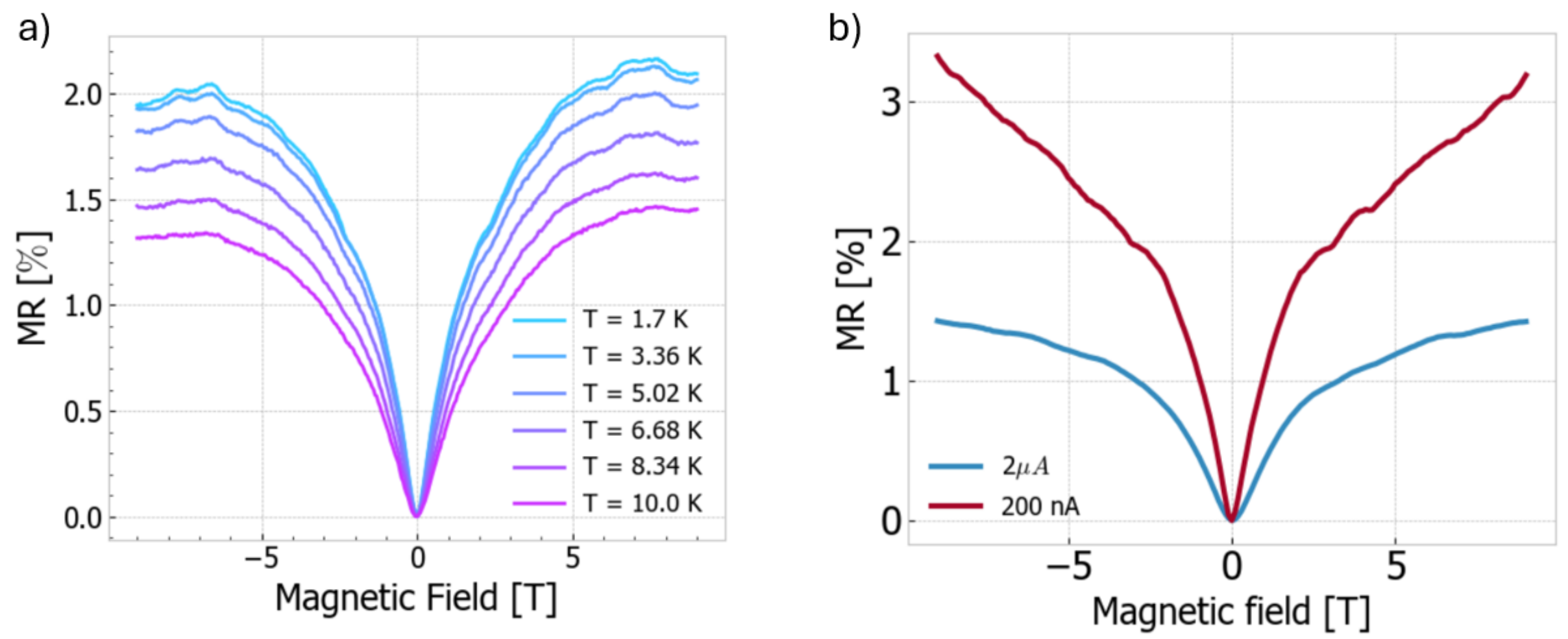

*Figure S4.1 Quantum Oscillation Dependencies. a) temperature dependence of the magnetoresistance vs. magnetic field plot, indicating higher oscillation amplitudes for lower temperatures. b) magnetoresistance for two different currents flowing through the device, showing self-heating effects for high currents.*

To extract the quantum oscillations from the magnetoresistance data, a fourth-order polynomial is fitted to the measured curve, as shown in Figure S4.2(a). The background fit is subtracted to isolate the oscillatory component, ΔR, which is then plotted against 1/B (Figure S4.2(b)). Since Shubnikov–de Haas oscillations are periodic in 1/B, their frequency is extracted using a Fast Fourier Transform (FFT), with the resulting power spectral density shown in Figure S4.2(c). Carrier concentration can be estimated using different methods. One common approach applies the Onsager relation, which links the oscillation frequency to the Fermi surface cross-sectional area. Assuming a spherical Fermi surface, this yields a carrier concentration of $n_F=1.91\cdot10^{17}$ $cm^{-3}$ at 1.7 K. However, the variation in frequency with magnetic field angle indicates an anisotropic Fermi surface, invalidating the spherical approximation. Alternatively, the carrier concentration can be estimated from the spacing between successive minima in the 1/B plot, resulting in $n=1.15\cdot10^{16}$ $cm^{-3}$ at 1.7 K.

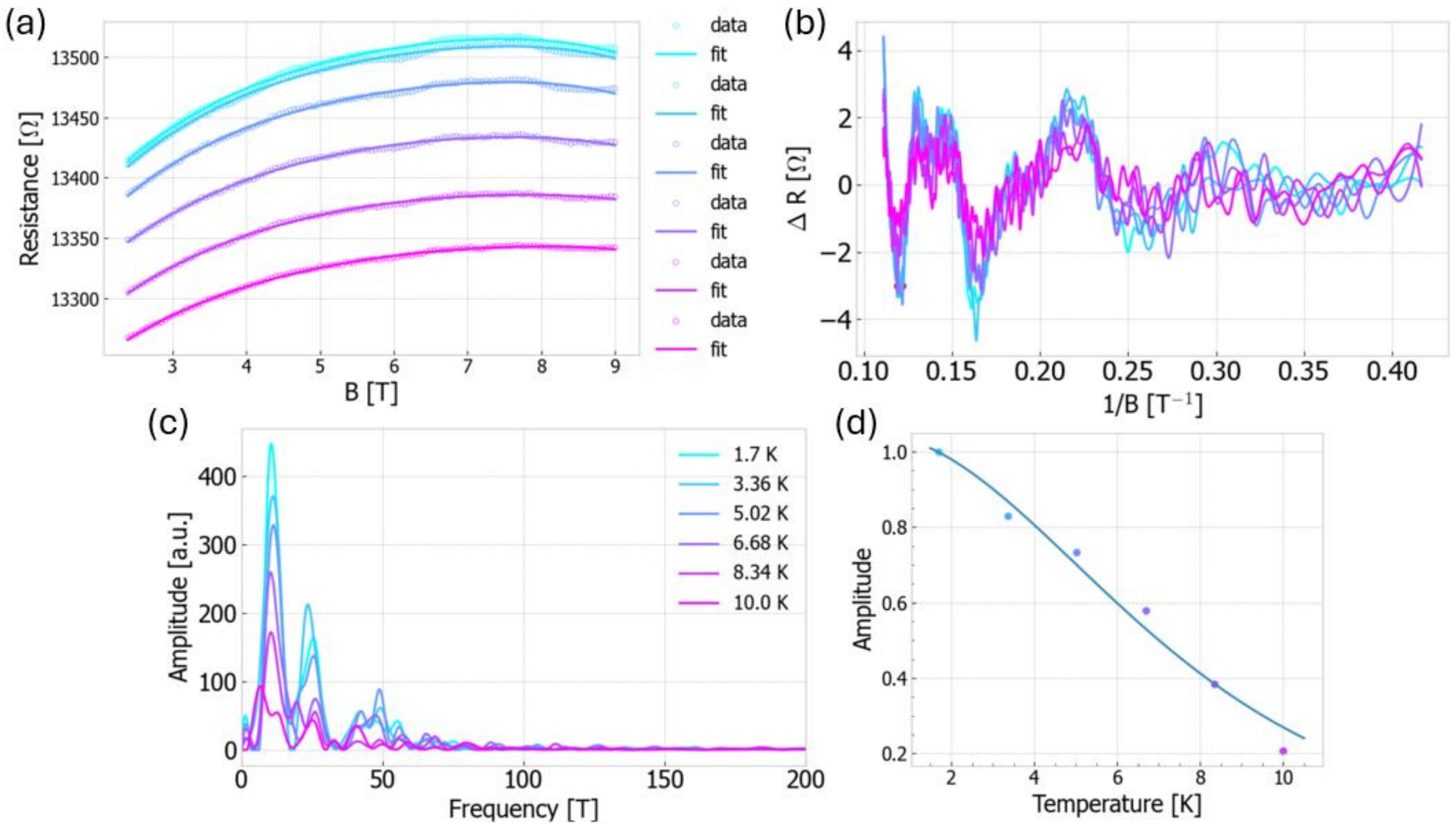

*Figure S4.2 Shubnikov-de-Haas data. a) measured and fitted data of the resistance vs. magnetic field measurements for different temperatures. b) subtraction of the measured data from the fit vs the magnetic field shows the quantum oscillations. c) shows the FFT of the oscillations d) the amplitude of the FFT vs. temperature.*

Further insight into the material properties is gained by analyzing the temperature dependence of the oscillation amplitude using Lifshitz–Kosevich theory. The normalized amplitude as a function of temperature is shown in Figure 11(d), from which an effective mass of $m^*=0.12m_e$ is extracted. Assuming a spherical Fermi surface, this corresponds to a mobility of $\mu_F$=1717 cm$^2$/Vs and a scattering time of $\tau_F$=1.18 ps. Using the lower carrier concentration, the mobility and scattering time increase to μ=28557 cm$^2$/Vs and τ=19.6 ns, respectively. While Shubnikov–de Haas oscillations offer valuable insights into transport properties, their application to nanowires is challenging due to the small number of detectable oscillations up to 9T and small signal amplitudes and high noise levels, resulting in significant uncertainty.

## S5: Noise level

To assess the noise level, a constant voltage is applied while monitoring the current through the nanowire over time. Figure S5.1(a) shows resistance fluctuations over 100 seconds for Sample A, with a random variation of about 1.75%. As resistive components inherently generate noise, so minimizing the contact resistance is important, and is done by annealing the contacts. After annealing (Figure S5.1 (c)), the fluctuation decreases to 0.6%. Surface effects are another likely noise source. To explore this, a Sample A device with annealed contacts was encapsulated in a 20 nm $Al_2O_3$ layer (Figure S5.1 (c)). This reduced the resistance variation further to just 0.07%, demonstrating the shell's effectiveness in suppressing electronic noise. The strong dependence on surface treatment suggests that surface conduction significantly contributes to overall transport in the nanowires.

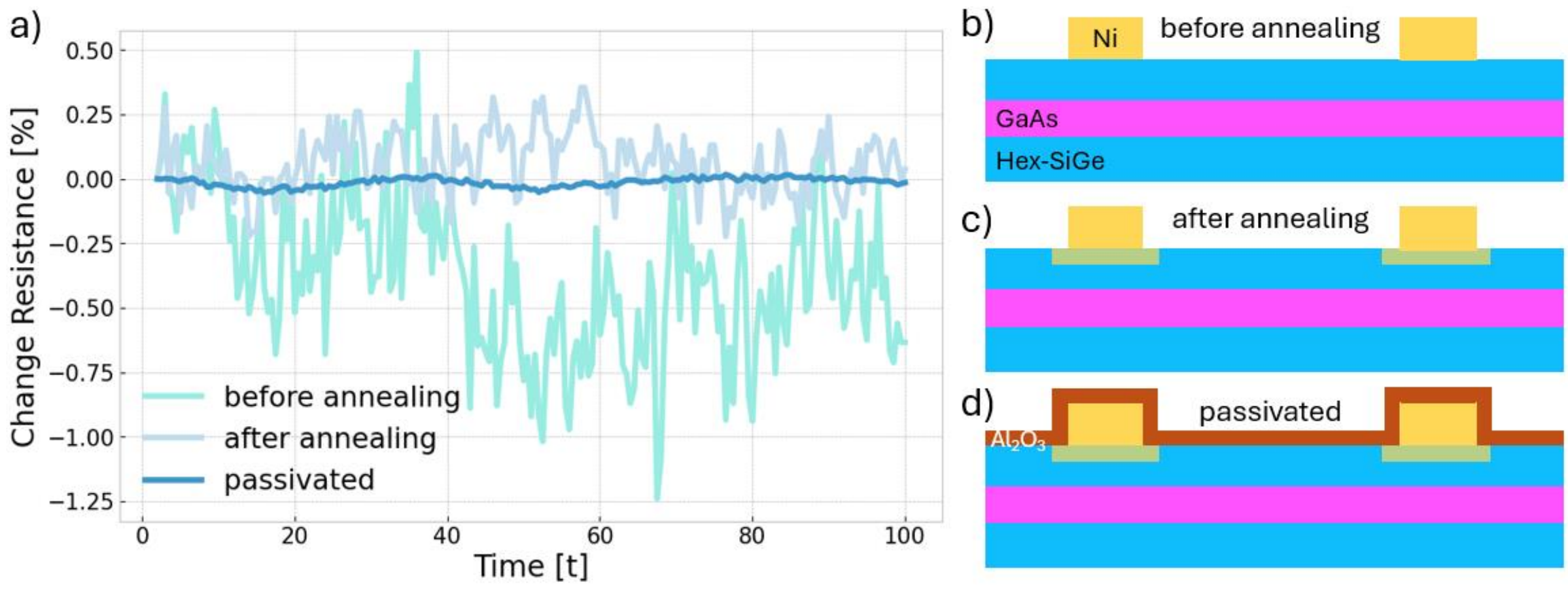

*Figure S5.1 Noise level Electrical noise and schematic illustration of hex-SiGe NW devices. a) the noise level in Sample A with contacts before and after annealing and for a surface passivated device. Schematic illustration of the device with contacts b) before annealing, c) after annealing, and d) passivated.*

## S6: Implantation and recrystallization

Figure S6.1 shows STEM images of the cross section of a nanowire after implantation doping, before (a) and after (b) recrystallization. A clear contrast is observed between the amorphous upper layer and the underlying crystalline region. As seen in (a), the thickness of the amorphous layer varies with the implantation angle relative to the sample surface, though it remains consistent (~35 nm) when measured along the implantation direction. This matches the expected value, given the peak doping concentration was targeted 20 nm below the surface. A bright line at the nanowire/Pt interface is also visible, likely indicating a native oxide layer formed upon air exposure. Additionally, sputtering damage from the implantation causes noticeable edge rounding. Figure 6.1 (a) presents a HRTEM image of the transition zone, where the lower darker region displays a crystalline lattice, while the lighter upper region lacks such structure, confirming amorphization.

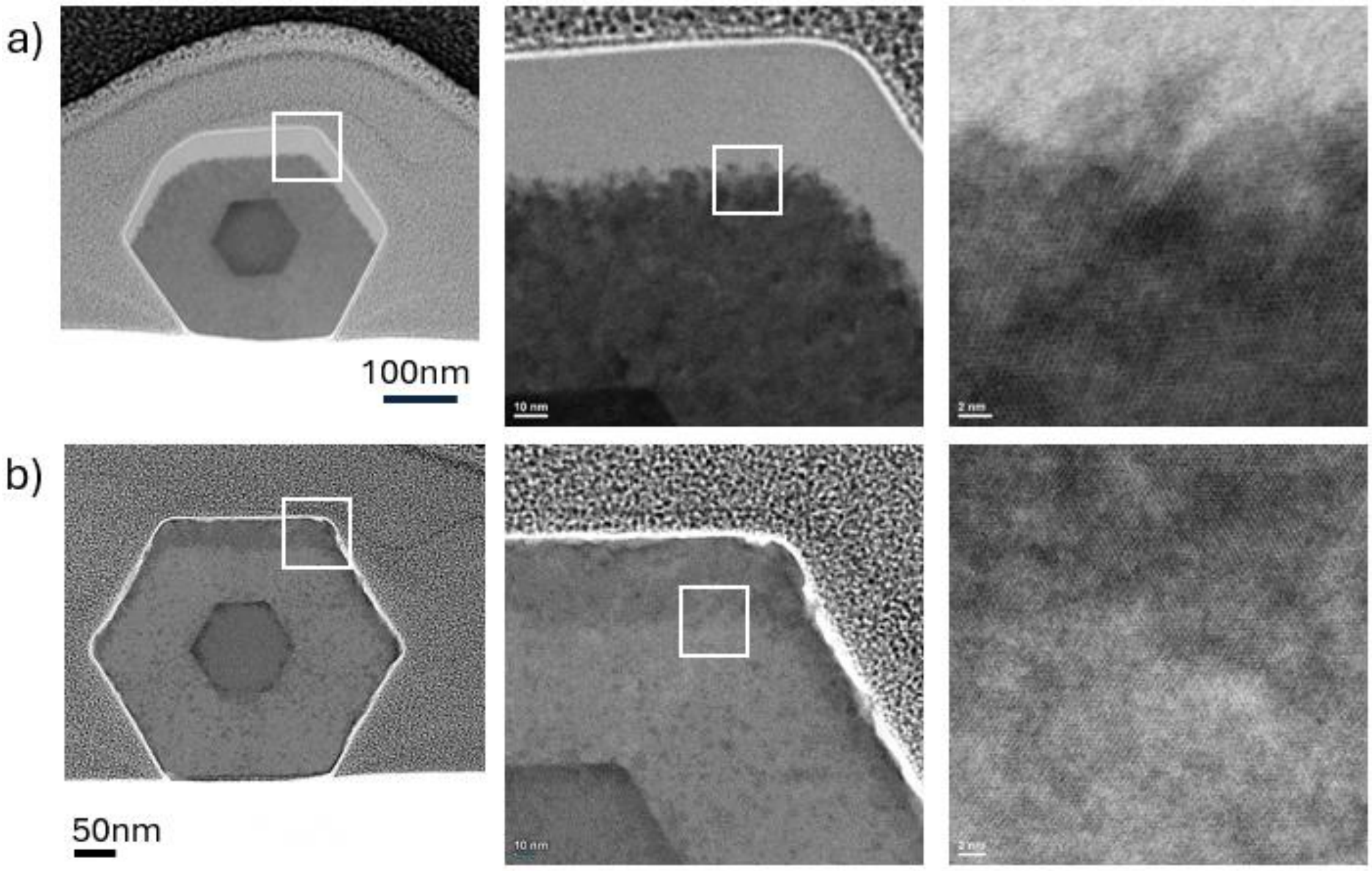

*Figure S6.1 Cross-sectional STEM in the [0001]-direction after implantation doping. a) The amorphous area is visible as uniform and a bright band. The amorphous layer is approximately 35 nm deep. The HRTEM image shows the crystalline (bottom) to amorphous (top) transition. b) STEM image the core/shell structure of the hex-GaAs/SiGe NW after recrystallization. The magnified STEM image on the edge of the NW and the implanted region. The HRTEM with atomic resolution of the interface between the original and the recrystallized lattice shows recrystallization.*

Recrystallization is crucial for restoring the hexagonal crystal structure. To achieve this, samples were annealed at 600 °C for 30 seconds. During this process, the underlying crystalline region acts as a seed for recrystallization of the amorphous layer. STEM cross-sections of a GaAs/SiGe core/shell nanowire after annealing (Figure S6.1 (b)) reveal that the implanted region remains visible as a darker contrast, suggesting residual lattice distortions. At the nanowire's edge, a thicker oxide layer is observed compared to the pre-annealing state, likely due to enhanced oxidation at high temperatures.

## S7: Richardson Plot Schottky Diode

Temperature-dependent IV-characteristics of a single Schottky diode can be used to extract key parameters such as the saturation current $I_0$, ideality factor n, and Schottky barrier height, using

$$\ln\left(\frac{1}{\left(1-\exp\left(\frac{-q(V-R_S I)}{kT}\right)\right)}\right)=\ln(I_0)+\frac{qV}{kT}$$

The corresponding data is plotted in Figure S7 (a). The equation shows that the y-intercept of a linear fit yields $I_0$, while the inverse slope is used to calculate the ideality factor at different temperatures (Figure S7 (c)). Varying the assumed series resistance $R_S$ (0 kΩ, 15 kΩ, 30 kΩ) significantly impacts the extracted ideality factor. The saturation current $I_0$ is further used to construct the Richardson plot (Figure S7 (b)), where the y-intercept provides the effective Richardson constant A* for a given junction area A. The Schottky barrier height $\phi_B$ is then calculated using

$$I = I_0 \exp\left(\frac{qV}{nkT}\right)\left(1 - \exp\left(\frac{-q\,V}{k\,T}\right)\right) \text{ with } I_0 = AA^{*}T^{2}\exp\left(-\frac{\phi_B}{kT}\right)$$

and shown in Figure S7(d). Both the barrier height and ideality factor depend strongly on the assumed series resistance; a value of 0 kΩ is unphysical and yields unrealistically small parameters, while a more plausible value of 15 kΩ produces consistent and realistic results.

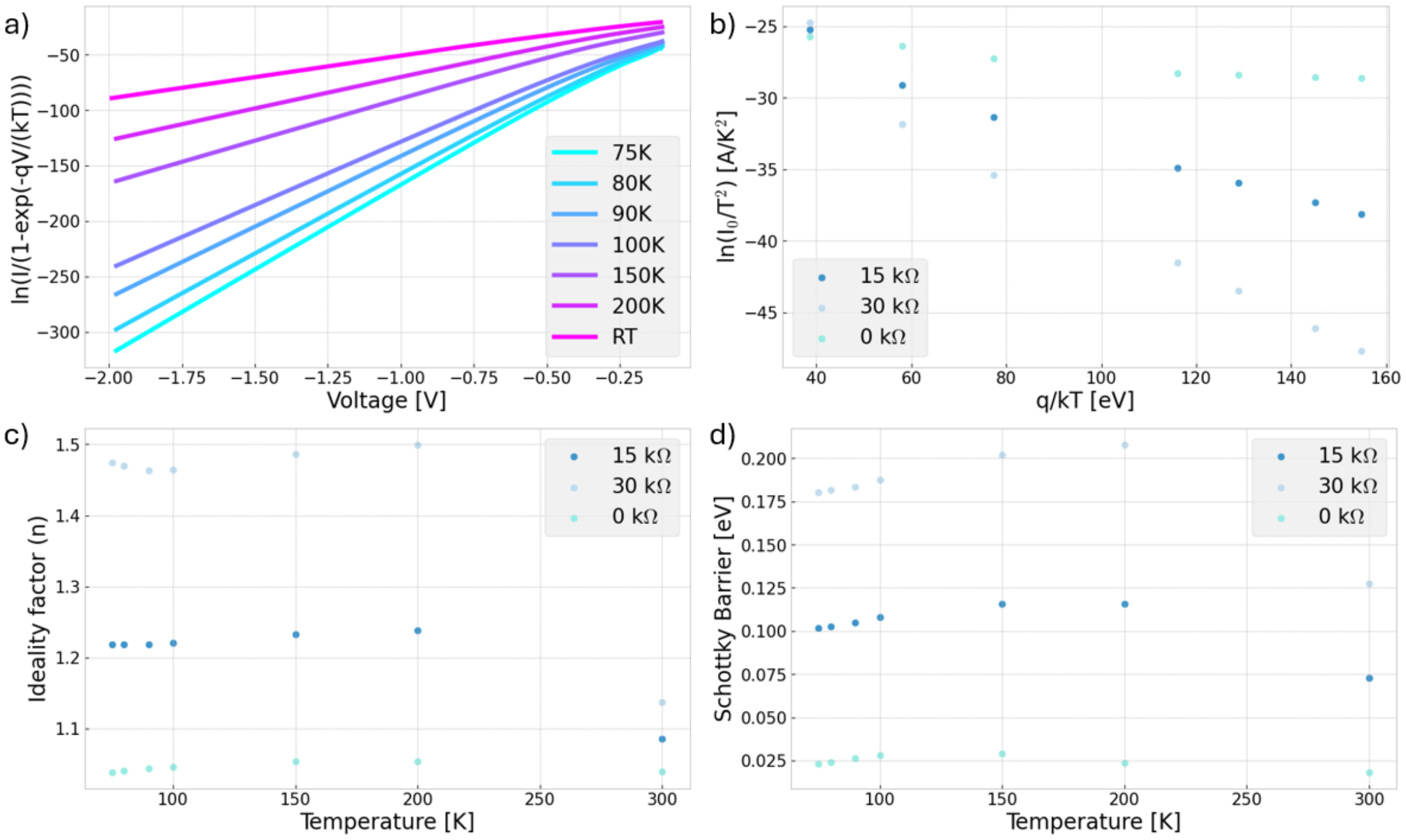

*Figure S7 Richardson plot a) ln(I) vs. V plot of the Schottky contact at various temperatures. b) Richardson plot with the reverse saturation current ($I_0$) calculated from the linear form of the ln(I) vs. V plot for different series resistances (0 kΩ, 15 kΩ, 30 kΩ). c) Temperature dependent ideality factor (n). d) Temperature dependent Schottky barrier potential.*